\documentclass[aps,prd,twocolumn,floatfix,nofootinbib]{revtex4}

\usepackage{graphicx}
\usepackage{epic}
\usepackage{eepic}
\usepackage{latexsym}
\usepackage{amsmath}

\newcommand{\eq}[1]{(\ref{#1})}
\newcommand{\be}{\begin{equation}}
\newcommand{\ee}{\end{equation}}
\newcommand{\bea}{\begin{eqnarray}}
\newcommand{\eea}{\end{eqnarray}}

\newcommand{\vs}[1]{\vspace{#1 mm}}
\newcommand{\hs}[1]{\hspace{#1 mm}}

\def\h{\hat}

\def\a{\alpha}

\def\cc{\gamma}

\def\d{\delta}
\def\D{\Delta}
\def\e{\epsilon}

\def\fr{\frac}
\def\F{\Phi}

\def\h{\eta}

\def\l{\lambda}

\def\m{\mu}
\def\n{\nu}

\def\th{\theta}

\def\z{\zeta}

\def\y{\eta}
\def\o{\omega}

\def\del{\partial}

\let\bm=\bibitem
\def\nn{\nonumber}

\begin{document}

\title{The imprint of primordial gravitational waves on the CMB intensity profile}
\author{Ali Kaya}
\affiliation{\vs{3}Department of Physics, Astronomy and Geophysics, Connecticut College, New London, CT 06320, U.S.A. \vs{5}}

\begin{abstract}
	
We use the induced geometry on the two dimensional transverse cross section of a photon beam propagating on a perturbed Friedmann-Robertson-Walker (FRW) spacetime to find the Cosmic Microwave Background (CMB) photon distribution over a telescope's collecting area today. It turns out that at each line of sight the photons are diluted along a transverse direction due to gravitational shearing. The effect can be characterized by two spin-weight-two variables, which are reminiscent of the Stokes polarization parameters. Similar to that case, one can construct a scalar and a pseudo-scalar function where the latter only gets contributions from the tensor modes. We analytically determine the power spectrum of the pseudo-scalar at superhorizon scales in a simple inflationary model and briefly discuss possible observational consequences. 

\end{abstract}

\maketitle

\section{Introduction}

After they decouple from the plasma at recombination, the CMB photons move freely along the null geodesics of the curved spacetime. Obviously, the exact geodesic lines are slightly modified by cosmological perturbations as compared to the unperturbed geodesics. In general, this small change can be decomposed into a component along the line of sight causing a redshift in the frequency and a displacement perpendicular to the line of sight yielding lensing effects. The gravitational lensing of  CMB photons is well studied, for a review see \cite{glcmb}, and the effect has also precise observational signatures \cite{ob1,ob2}. 

Previous studies mostly focus on how the CMB temperature map on the sky is modified by the lensed angular positions of the CMB photons. The deflection angle caused by lensing becomes a pure gradient and the corresponding  potential is introduced as the main statistical variable. There is also some work on the lensing shear effect that modifies the so called hot and cold spot CMB ellipticity distribution \cite{el1,el2,el3}. Since the background temperature map is uniform, these are nonlinear effects that appear at the second order. Moreover, they are also dominated by the density perturbations; the influence of the tensor modes is completely negligible. Yet, it is also possible to extract a rotational component of the shear which have contributions only from the gravitational waves \cite{gw0,gw1} (this is similar to the  $B$-mode of the CMB polarization) but unfortunately the presumed signal is very small and below the noise level \cite{gw1,gw2,gw3}. 

The gravitational lensing effects can also be studied using the geodesic deviation equation. In that framework, one calculates the expansion, shear and rotation parameters as the basic geometrical variables of the congruence. In a recent work \cite{ak}, we have instead determined the induced two dimensional metric on the transverse cross section of a null geodesic beam in a perturbed FRW background. We have shown that the transverse metric does not depend on the slicing and its derivative along the geodesic flow can be decomposed to yield the expansion, shear and rotation. Clearly, the induced metric offers a direct geometrical description of a photon congruence.  

Consider the evolution of a CMB photon beam back in time from the moment of its capture by a telescope today to the time of decoupling. The transverse slice corresponding to the telescope collection area is mapped to another slice of the beam at decoupling. The distribution of the trajectories on the initial surface is expected to be uniform since the photons are in local thermal equilibrium. However, the distribution of the photons hitting the telescope surface today would in general be nonuniform because of the gravitational shearing. Each photon trajectory marks a point on a transverse slice of the beam and we call the distribution of these points {\it the intensity profile} of the congruence. In \cite{ak} we have shown that the CMB intensity profile is characterized by two variables that are reminiscent of the Stokes polarization parameters. In this work, we will elaborate more on these variables; specifically we will construct a pseudo scalar quantity which is only generated by the primordial gravitational waves, similar to the $B$-mode of polarization. 

\section{Null geodesics on perturbed FRW} 

Let us look at the null geodesics on the following perturbed FRW spacetime 
\be\label{met}
ds^2=a(\h)^2\left[-(1+2\Psi)d\h^2+[(1+2\Phi)\d_{ij}+\cc_{ij}]dx^i dx^j\right]. 
\ee
By definition the tensor mode is traceless $\d^{ij}\cc_{ij}=0$ and at the moment we do not impose any gauge fixing conditions. In the present study, we will only work in the linear theory, therefore {\em all equations written below must be assumed to be valid up to the first order in perturbations.} 

To determine the null geodesic trajectories, one can first solve for the {\it tangent vector field} on the spacetime obeying
\be
 p^\m\nabla_\m \,p^\n=0, \hs{7} p^\m p_\m=0.
\ee
Defining the perturbations around the unperturbed field by
\bea
&&p^0=\fr{1}{a^2}+\d p^0,\nn \\
&&p^i=\fr{l^i}{a^2}+\d p^i,\hs{5}\d_{ij}l^i l^j=l^il^i=1,\label{g} 
\eea
one can fix $\d p^0$ from $p^\m p_\m=0$ as
\be\label{dp0}
\d p^0=l^i\d p^i+\fr{1}{a^2}\left(\Phi-\Psi\right)+\fr{1}{2a^2}\cc_{ij}l^i l^j
\ee
and solve for $\d p^i$ so that 
\bea
&&\d p^i(x,\h)=-\fr{2}{a(\h)^2}l^i\,\Phi(x,\h)-\fr{1}{a(\h)^2}\cc_{ij}(x,\h)\,l^j \nn\\
&&+\fr{1}{a(\h)^2}\del_i\int_{\h_0}^{\h} d\h'\left[\Phi- \Psi+\fr12 \cc_{jk}l^j l^k\right]\left(x_{\h'\h},\h'\right)\nn\\
&&+\fr{1}{a(\h)^2}\left[2l^i\,\Phi+\cc_{ij}\,l^j\right](x_{\h_0\h},\h_0)\label{dpi}
\eea
where $\h_0$ is the present conformal time and we introduce $x^i_{\h_1\h_2}$ to be the spatial position on an unperturbed geodesic path
\be\label{xl}
x_{\h_1\h_2}^i=x^i+l^i(\h_1-\h_2).
\ee
This is the unique solution obeying the condition 
\be\label{ic0}
\d p^i(x,\h_0)=0
\ee
hence $p^i(x,\h_0)=l^i/a(\h_0)^2=l^i$, where we also set $a(\h_0)=1$. As a result, $l^i$ defines the present direction of propagation and the actual line of sight including the lensing effects. Note that the time argument of the fields in the last line in \eq{dpi} is the present time $\h_0$. These terms arise since we demand \eq{ic0} and they vanish when the derivative operator along the unperturbed geodesic $\del_\h+l^i\del_i$ is applied. 

These equations are worked out for a photon having ``unit" energy and the general case can be obtained by scaling $p^\m\to E p^\m$. Our results below do not depend on the parameter $E$ and therefore we are not going to introduce it. 

It is possible to obtain the Sachs-Wolfe effect using the above equations. The 4-velocity vector of a comoving observer in \eq{met} (obeying $u^\m u_\m=-1$) can be found as
\be\label{u} 
u^0=\fr{1}{a}(1-\Psi),\hs{5}u^i=0
\ee
and the energy of a photon as measured by this observer is given by 
\be\label{up}
\o=-u^\m p_\m. 
\ee
One can see that 
\be\label{w}
\o=p=\fr{1}{a}\left[1+a^2 l^i\d p^i+\F+\fr12\cc_{ij}l^i l^j\right],
\ee
where 
\be
p=\sqrt{g_{ij}p^i p^j }.
\ee
As it was first observed in \cite{rd}, \eq{w} encodes the Sachs-Wolfe effect if one defines $T\propto \o$. Indeed, by applying the derivative along the geodesic trajectory $p^\m\del_\m$ one can see   
\be\label{sw}
\left(\fr{\del}{\del \h}+l^j\del_j\right)\fr{\d T}{T}=-\Phi'-l^i\del_i\Psi-\fr12 \cc_{ij}'l^il^j,
\ee
which exactly gives the evolution of the temperature fluctuations along the unperturbed geodesic lines. In \cite{ak} we have shown that \eq{sw} is valid for any (and not necessarily thermal) distribution function provided one reads the temperature from the average intensity by $T^4\propto I$. 

One can obtain the geodesic path $x^\m(\l)$ by integrating 
\be\label{tv}
\fr{dx^\m}{d\l}=p^\m(x(\l)),
\ee
where $\l$ is an affine parameter. The zeroth component of the above equation can be used to relate $\l$ and the conformal time as
\be
\fr{d}{d\l}=\left(\fr{1}{a^2}+\d p^0\right)\fr{d}{d\h}.
\ee
Defining the perturbed geodesic 
\be\label{gx}
x^i(\h)=x_0^i +l^i(\h-\h_0)+\d x^i(\h),
\ee
\eq{tv} implies
\be
\fr{d\d x^i}{d\h}=a^2\d p^i - a^2l^i\d p ^0.
\ee
After using \eq{dp0}, one can integrate to obtain
\bea
&&\d x^i(\h)=\left[\d^{ij}-l^i l^j\right]\int_{\h_0}^\h d\h'a^2(\h')\,\d p^j(x_{0\h'\h_0},\h')\nn\\
&&\hs{3}+l^i\int_{\h_0}^\h d\h'\left[\Psi-\Phi-\fr12\cc_{jk}l^j l^k\right](x_{0\h'\h_0},\h'), \label{dxi}
\eea
where $\d p^j$ is found in \eq{dpi} and the spatial argument of the functions $x_{0\h'\h_0}$ stands for $x^i_{0\h'\h_0}=x^i_0+l^i(\h'-\h_0)$,
as in \eq{xl}. Note that \eq{dxi} obeys $\d x^i(\h_0)=0$ and thus \eq{gx} yields the unique null geodesic path which passes from the spatial position $x^i_0$ at time $\h_0$ along the direction $l^i$. 

\section{The geometry of the photon beam and gravitational shearing} 

Eq. \eq{gx}, where $\d x^i$ is given in \eq{dxi}, actually describes a family of geodesics parametrized by the constants $x^i_0$ and $l^i$. A photon beam observed at $\h_0$ along direction $l^i$ corresponds to a (small) subset in that family. Let $\D x_0^i$ denote the coordinate difference between two nearby geodesic lines at $\h_0$. The time evolution of this interval can be found from the solution \eq{gx} 
\be\label{tev} 
\D x^i(\h)=\D x_0^i +\D \d x^i(\h),
\ee
where $\D \d x^i(\h)$ is obtained by varying \eq{dxi} with respect $x^i_0$. The corresponding physical length is given by the metric 
\be\label{tmet1}
\left|\D x^i(\h)\right|=\left[g_{ij}(\h)\D x^i(\h)\D x^j(\h)\right]^{1/2}.
\ee
At the time of observation, the transverse cross section of the beam can be specified by the vectors $m^i$ and $n^i$, where  $(l^i,m^i,n^i)$ forms an orthonormal set with respect to $\d_{ij}$ (the impact of the metric perturbations at that instant is completely negligible). The evolution of the transverse beam cross section can be found by choosing $\D x_0^i=L m^i$ or $\D x_0^i=L n^i$ in \eq{tev}, where $L$ is the telescope size. We have checked that the two dimensional metric obtained from \eq{tmet1} (involving the displacements $\D x_0^i=L m^i$ and $\D x_0^i=L n^i$) exactly agrees with the slicing independent transverse metric obtained from the geodesic deviation equation in \cite{ak} ($l^i$ had been chosen as the geodesic cotangent at the time of decoupling in \cite{ak}).

\begin{figure}
	\centerline{\includegraphics[width=9cm]{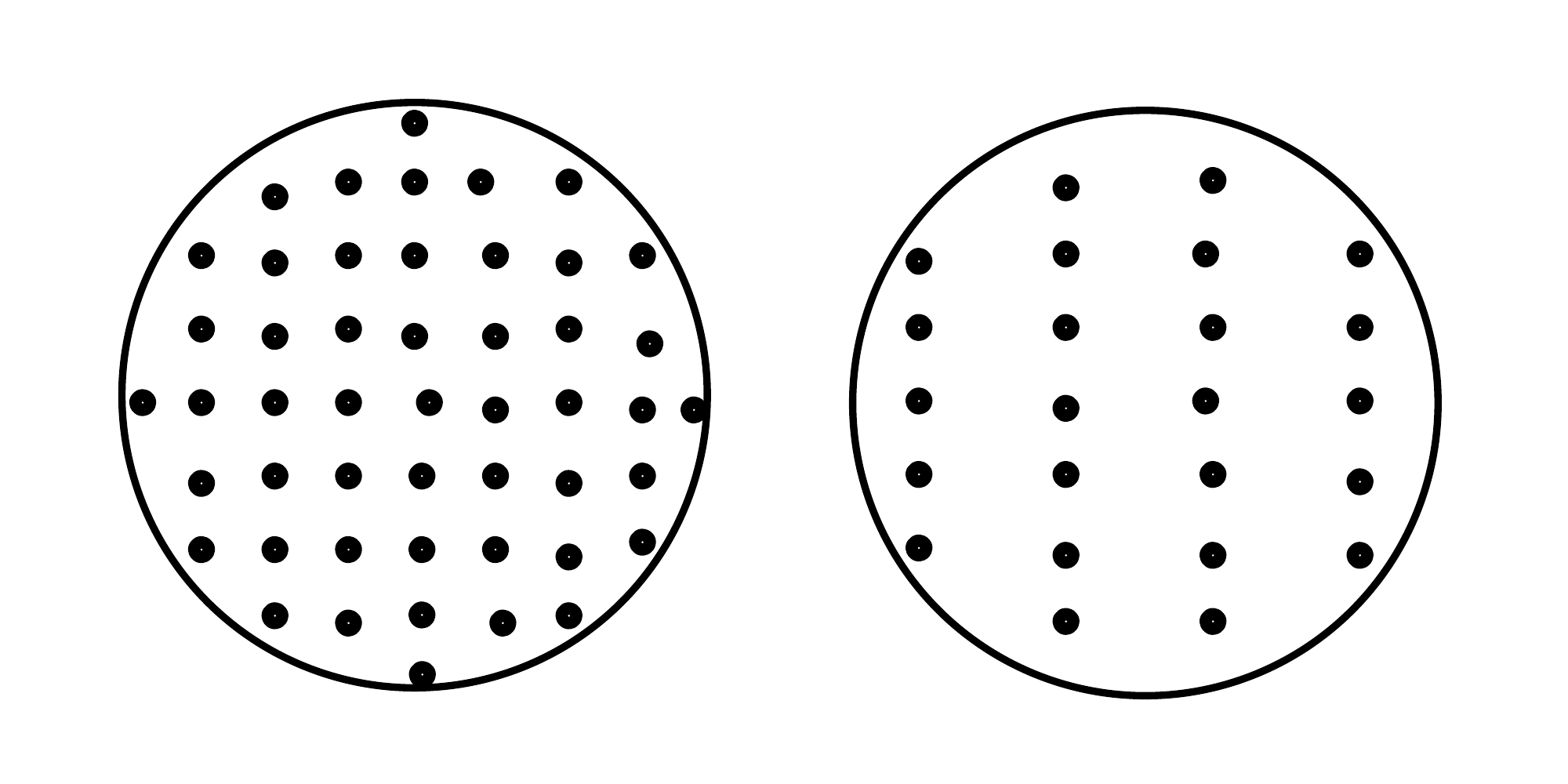}}
	\caption{CMB photons hitting a detector. On the left there is a uniform distribution over the area but on the right the photons are  diluted along the $x$-axis as compared to the $y$-axis due to gravitational lensing.} 
	\label{fig1}
\end{figure}

We now compare the physical lengths of the two transverse directions $Lm^i$ and $Ln^i$ at the time of recombination $\h_r$. Their difference equals $La(\h_r)Q$, where we define 
\be\label{uu}
Q=\left[\fr12 \cc_{ij}(x_{0\h_r\h_0},\h_r)+\del_{x^j_0}\d x^i(\h_r)\right](m^i m^j-n^i n^j).
\ee
One can also determine the size difference between $\pi/4$ rotated directions $(m^i+n^i)/\sqrt{2}$ and $(m^i-n^i)/\sqrt{2}$, which can be found as $La(\h_r)U$, where  
\be\label{vv}
U=\left[\fr12 \cc_{ij}(x_{0\h_r\h_0},\h_r)+\del_{x^j_0}\d x^i(\h_r)\right](m^i n^j+n^i m^j).
\ee
The two parameters $Q$ and $U$, which depend on the directions $(l^i,m^i,n^i)$, identify the shape of the initial transverse surface  at the time of recombination, {\it which has a uniform photon distribution over it.} Obviously, while this initial surface evolves to become the (circular) cross section today, the photons are diluted in the direction that expands more compared to the other, see Fig \ref{fig1}. 

\begin{figure}
	\centerline{\includegraphics[width=9cm]{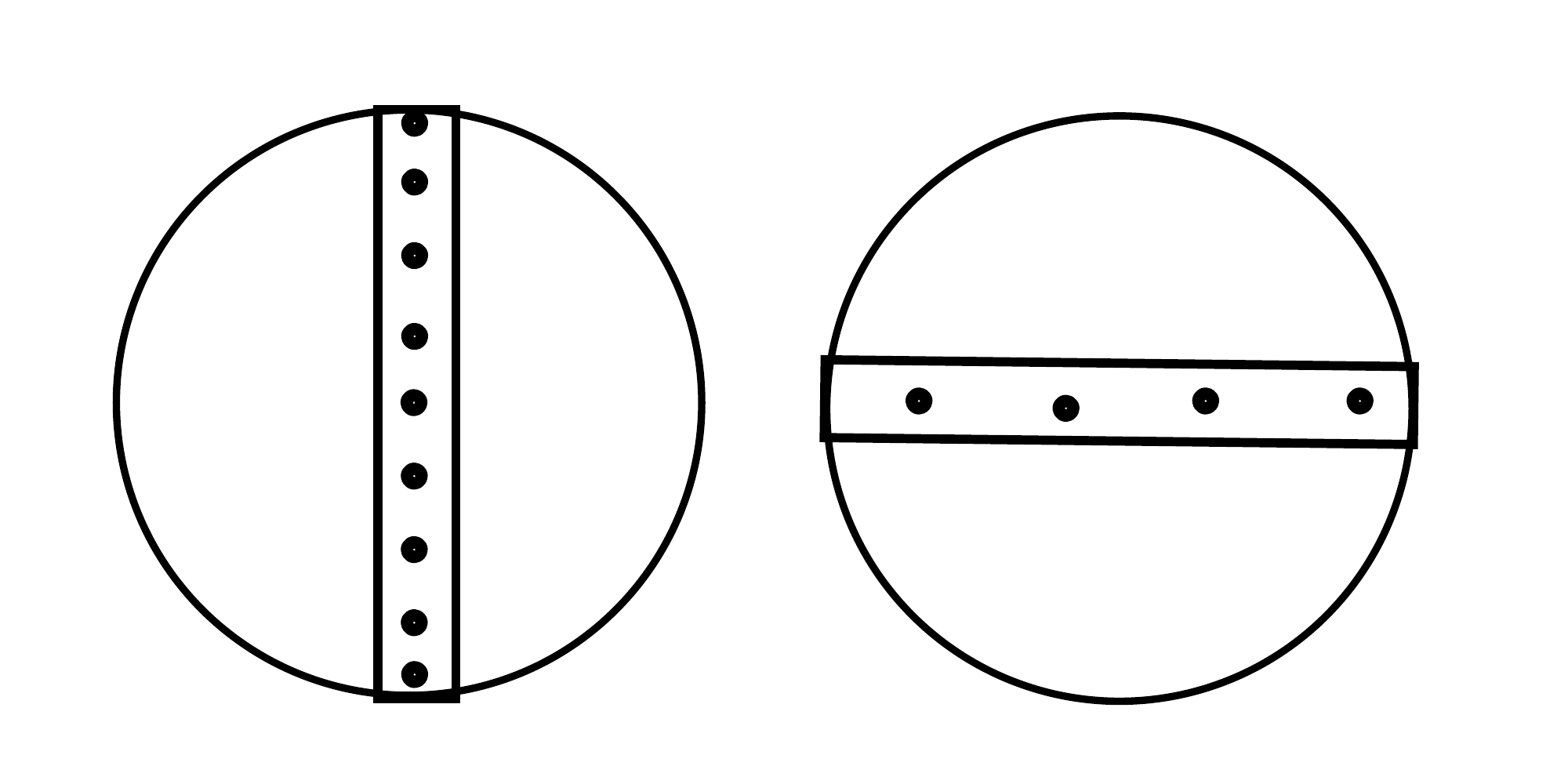}}
	\caption{The lensed photons impinging on two narrow slits instead of the whole detector area. The observed ``intensities" are not equal because the number of incident photons is different for each case.} 
	\label{fig2}
\end{figure}

The phase space volume element along a geodesic flow does not change by Liouville's theorem and this leads to the standard rule that gravitational lensing does not modify specific intensity, see e.g. \cite{mtw}. For the variables $Q$ and $U$, this result is avoided since these do not directly measure the intensity; instead they are related to the distribution of photons over a transverse surface (which we call the intensity profile of the beam). Obviously, the validity of the particle description is crucial for the observability of this effect. Relying on the photon picture, one can quantify the surface distribution by measuring the energy flux over narrow slits instead of the whole area. For a given wavelength, the slit width must be small enough so that the usual concept of intensity fails (note that intensity is a coarse grained concept in the photon picture). In that case, $Q$ becomes proportional to the energy flux difference between two slits extending along $m^i$ and $n^i$ directions, see Fig. \ref{fig2}. Likewise, the flux difference between $\pi/4$ rotated slits gives $U$. 

One can simplify \eq{uu} and \eq{vv} to a very good approximation by computing the leading order contributions. From \eq{dxi}, the terms coming from $\d x^i(\h_r)$ can be seen to appear inside single or double time integrals, which give oscillating contributions. The tensor modes in these integrals are negligible compared to the first term in \eq{uu} and \eq{vv}. Using also $\Psi\simeq -\Phi$ and ignoring the monopole term, one can obtain 
\bea
&&Q\simeq\left[\fr12 \cc_{ij}(x_{0\h_r\h_0},\h_r)+K_{ij}\right](m^i m^j-n^i n^j)\nn \\
&&U\simeq\left[\cc_{ij}(x_{0\h_r\h_0},\h_r)+2K_{ij}\right]m^i n^j,\label{uva}
\eea
where
\be\label{kij}
K_{ij}=2\fr{\del^2}{\del x_0^i \del x_0^j}\int_{\h_0}^{\h_r}d\h'(\h_r-\h')\Phi(x_{0\h'\h_0},\h').
\ee
The explicit $(\h_r-\h')$ factor in $K_{ij}$ appears after changing the order of a double time integral. 

In \eq{uva} the scalar mode contributions involve an oscillating time integral but they are still expected to dominate the power spectra over gravitational waves. Therefore, it is desirable to construct a variable which only depends on the tensor modes. The doublet $(Q,U)$ rotates by $2\a$ when the tangent vectors $(m,n)$ are rotated by $\a$. Thus they constitute spin-weight-two objects on the sphere. The infinitesimal variations of $l^i$ (that respect the constraint $l^i l^i=1$) can be parametrized like
\be
\d l^i= (\d a) m^i + (\d b) n^i,
\ee
where the doublet $(\d a,\d b)$ has spin-weight $-1$. The derivative operator $(\d^2/\d a^2-\d^2/\d b^2,2\d^2/\d a\d b)$ has spin-weight 2 and by applying it on $(Q,U)$ with Kronecker delta and epsilon tensor contractions, one can obtain spin-weight-zero scalar and pseudo-scalar on the sphere 
\bea
&&E(l)=\left[\fr{\d^2}{\d a^2}-\fr{\d^2}{\d b^2}\right]Q+\fr{2\d^2U}{\d a\d b},\nn\\
&&B(l)=\fr{2\d^2Q}{\d a\d b}-\left[\fr{\d^2}{\d a^2}-\fr{\d^2}{\d b^2}\right]U. \label{EB1} 
\eea
It is easy to see that $\Phi$ drops out in $B$ which indeed becomes 
\bea
B(l)&=&(\h_0-\h_r)^2\,\fr{\del^2 \cc_{ij}(x_{0\h_r\h_0},\h_r)}{\del x^k_0\, \del x^l_0}\\
&&\left[m^k n^l(m^i m^j-n^i n^j)-m^in^j(m^km^l-n^kn^l)\right].\nn
\eea
Just like in the polarization case, $E$ and $B$ represent curl and divergence free field lines, this time, formed by the ``eigen-directions" of $(Q,U)$ on the sphere (the eigen-direction at a given point can be defined from one of the vectors of the basis $(m,n)$ in which $(Q,U)$ becomes proportional to $(1,0)$, i.e. one has $U=0$).

\section{The power spectra} 

We work out the power spectra of these variables at superhorizon scales in a simplified model having only two epochs, inflation and radiation. The scale factor in such a model is given by 
\be\label{mct}
a(\y)=\begin{cases}-\fr{1}{H_I\y}&\y\leq \y_I,\vs{3} \\ 
	H_0(\y-2\y_I)&\y_I\leq \y,
\end{cases}
\ee
where $\y_I=-1/\sqrt{H_IH_0}$, and $H_I$ and $H_0$ are the Hubble parameters at inflation and today, respectively. The form of \eq{mct}  is fixed by demanding the continuity of the scale factor and the Hubble parameter at $\h_I$. The present conformal time can be found from $a(\h_0)=1$ which gives $\h_0=1/H_0+2\h_I$. The redshift at recombination is given by 
\be\label{zr}
z_r=\fr{\h_0}{\h_r}
\ee
and one may take $z_r\simeq 10^3$. Note the following hierarchy $\h_0\gg\h_r\gg|\h_I|$.

The mode function of a minimally coupled massless scalar field that is released in its Bunch-Davies vacuum at inflation is given by 
\be\label{bdv}
\m_k=\begin{cases}\fr{1}{\sqrt{2k}}\left[1-\fr{i}{k\y}\right]e^{-ik\y} & \y\leq\y_I,\vs{3}\\
	\m_k^I\cos\left[k(\y-\y_I)\right]+\fr{\m_k^I{}'}{k}\sin\left[k(\y-\y_I)\right] & \y_I<\h,	
\end{cases}	
\ee
where $\m_k^I=\m_k(\y_I)$, $\m_k^I{}'=\m_k'(\y_I)$ and prime denotes $\h$ derivative. 

The tensor perturbation can be expanded in terms of the mode functions  
\be
\cc_{ij}=\fr{1}{(2\pi)^{3/2}}\int d^3k\, e^{i\vec{k}.\vec{x}}\,\cc_k(\y)\, \e_{ij}^s \tilde{a}^s_{\vec{k}}+h.c.\nn
\ee
where $s=1,2$ and the creation-annihilation operators satisfy the usual commutation relations, e.g. $[a_k,a^\dagger_{k'}]=\d^3(k-k')$. The polarization tensor $\e^s_{ij}$ has the following properties
\bea
&&k^i\e^s_{ij}=0,\hs{5} e^s_{ii}=0,\hs{5}\e^s_{ij}e^{s'}_{ij}=2\d^{ss'}.\nn\\
&& \e^s_{ij}e^{s}_{kl}=P_{ik}P_{jl}+P_{il}P_{jk}-P_{ij}P_{kl},\label{tme}
\eea
where $P_{ij}=\d_{ij}-k^i k^j/k^2$. The tensor mode function $\cc_k(\y)$ can be determined in terms of $\m_k(\h)$ in \eq{bdv} as
\be\label{cck}
\cc_k=\fr{2}{a M_p}\m_k,
\ee
where $M_p$ is the reduced Planck mass $M_p^2=1/(8\pi G)$.

The gravitational potential $\Phi$ is determined from the curvature perturbation $\z$, which is conserved at super-horizon scales and can be expanded like \eq{tme}. The corresponding mode function during inflation can be taken as 
\be\label{zk}
\z_k=\fr{1}{a \sqrt{2\e} M_p}\m_k,
\ee
where $\e$ is the slow-roll parameter (for constant $\e$, $\z_k$ is actually given by the first Hankel function but \eq{zk} is a very good approximation when $\e\ll1$). The standard gauge fixing breaks down in reheating after inflation but there are alternative smooth gauges which would imply the standard results \cite{ak2}. The gravitational potential can be obtained by applying a coordinate change that sets the shift variable of the metric to zero, $N^i=0$. This yields
\be
\Phi_k=-\fr{\dot{H}a^2}{Hk^2}\dot{\z}_k,
\ee
where the dot denotes derivative with respect to the proper time $dt=ad\h$ and $H=\dot{a}/a$ is the Hubble parameter after inflation. 

In general, a two-point function involving the variables $Q$, $U$, $E$ and $B$ is specified by two distinct vector sets $(l_1,m_1,n_1)$ and $(l_2,m_2,n_2)$. One can conveniently choose $(l,m,n)=(\hat{r},\hat{\th},\hat{\phi})$ so that the angular integrals in the correlators become straightforward. The remaining (radial) momentum integrals contain the usual (distributional) UV infinities, which can be cured by $i\e$-terms (see \cite{ak3} for the implementation of the $i\e$-prescription in cosmology). In the following we take  
\be\label{th}
\th >\fr{1}{z_r},
\ee
where $\th$ is the angle between $l_1$ and $l_2$; i.e. $\cos(\th)=l_1^i l_2^i$. On the last scattering surface \eq{th} corresponds to superhorizon scales. 

Although the oscillating time integrals diminish their power, we estimate that the scalar modes still dominate the expectation values $\left<Q_1Q_2\right>$ and $\left<U_1U_2\right>$ (one has  $\left<Q_1U_2\right>=\left<Q_2U_1\right>=0$ identically) because of the slow-roll enhancement $1/\e$ coming from the curvature perturbation \eq{zk}. The angular integrals in momentum space give an oscillating factor which effectively sets a (comoving) cutoff scale for the remaining radial momentum integral (the cutoff is equivalent to the UV improvement implied by the $i\e$-prescription). This scale is roughly proportional to $\h_0$ and from \eq{kij}, which encodes the contribution of the scalar perturbation, one sees that on dimensional grounds while the two spatial derivatives yield $1/\h_0^2$ the time integrals give $\h_0^2$. This shows that in $\left<Q_1Q_2\right>$ and $\left<U_1U_2\right>$, the order of magnitude contributions of the tensors and the scalars are similar to the amplitudes of the expectation values $\left<\cc\cc\right>$ and $\left<\Phi\Phi\right>$, respectively. Hence the tensors are suppressed by the slow-roll parameter and only the $BB$-correlator is relevant for the gravitational waves. 

As usual in the two-point function $\left<B(l_1)B(l_2)\right>$ the oscillating subhorizon modes give negligible contributions when $\th$ obeys \eq{th}. Thus, to a very good approximation one can use the superhorizon spectrum (which can be obtained from \eq{cck} and \eq{bdv} when $k\h_r\ll1$)
\be\label{shs}
\left|\cc_k\right|^2\simeq\fr{2H_I^2}{M_p^2}\,\fr{1}{k^3}. 
\ee
In that case, the momentum integral can be calculated exactly without any issues (the integral is convergent at IR as $k\to0$ and its UV behavior is cured by the $i\e$-prescription). The result contains many terms when $i\e\not=0$, but in the limit $i\e\to0$ one gets a remarkably simple final formula
\be\label{bbs}
\left<B(l_1)B(l_2)\right>\simeq -\fr{2H_I^2}{\pi^2M_p^2}\cot^2(\th/2), \hs{5} \th >\fr{1}{z_r}.
\ee
Note that $B(l)$ is not a positive operator, hence a negative expectation value on scales \eq{th} is conceivable.

\section{Conclusions}

Eq. \eq{bbs} is the main result of this work. It gives a distinctive superhorizon signal that starts from zero at $\th=\pi$ and increases in magnitude with decreasing $\th$. Indeed, \eq{bbs} greatly enlarges as $\th$ approaches the subhorizon-superhorizon border (of this simple model) at $\th=1/z_r$. Of course, one would not expect \eq{bbs} to be correct up to that order since the subhorizon corrections to \eq{shs} become more and more important. 

The amplitude in \eq{bbs} depends directly on the scale of inflation, which is encoded by the Hubble parameter $H_I$. This is an expected feature for a power spectrum involving gravitational waves. Using the typical upper limit $H_I\simeq 10^{-5} M_p$, which can be obtained from the upper observational limit on the tensor-to-scalar ratio, one finds a very small amplitude, of the order of $10^{-10}$. Of course, this is the largest estimate since $H_I$ can be much smaller. Note that as they are defined, $Q$, $U$, $E$ and $B$ are all dimensionless variables and they measure relative magnitudes, e.g. if $I_V$ and $I_H$ are the intensities corresponding to the vertical and horizontal slabs in Fig. \ref{fig2}, then $Q=(I_V-I_H)/((I_V+I_H)/2)$. Therefore the figure $10^{-10}$ estimates a dimensionless signal (related to the variations of $Q$ and $U$ on the sphere), which can be compared to the usual temperature fluctuations having relative order of magnitude $10^{-5}$.

In any case, the result is encouraging for further investigations in a realistic model including small angles. Note that \eq{bbs} does not depend on the photon frequency and it can be determined from flux measurements as discussed above. These are technical advantages in terms of observability but detecting the corresponding signal will be hard if not impossible. Nevertheless, it is valuable to have an (even in principle) alternative to the CMB polarization experiments, as observing a quantum gravitational wave effect is already expected to be quite difficult. 

\begin{acknowledgments}
I am grateful for the support of IIE-SRF fellowship program and thank the colleagues at Connecticut College for their hospitality.
\end{acknowledgments}

\end{document}